\documentclass[a4paper]{article}

\usepackage[english]{babel}
\usepackage[utf8x]{inputenc}
\usepackage{amsmath}
\usepackage{graphicx}
\usepackage[colorinlistoftodos]{todonotes}

\title{Strong emergence in condensed matter physics}
\author{Barbara Drossel, Institute of Condensed Matter Physics, TU Darmstadt}

\begin{document}
\maketitle

\hskip 1cm \parbox{9cm}{\small  

} 

\medskip

\medskip

\medskip
\section{Introduction}

When I was a physics student, I often had the impression that I did not really understand the material that was presented. The reason was that the presented calculations were claimed to be based on the 'fundamental theory', usually quantum mechanics, but many of the steps that were made did not seem to really follow from the equations of quantum mechanics. These equations are deterministic, linear in the wave function, and invariant under reversal of the direction of time. In contrast, the calculations presented in the classes appeared to involve concepts that are incompatible with these features. Time reversal symmetry was broken when dealing for instance with the scattering of a quantum particle at a potential: the incoming particle is assumed to be not affected by the potential, but the outgoing particle is. Chance is introduced when basing all of statistical physics on probabilities, or when transition probabilities between quantum states are calculated, as for instance in scattering theory. The supposedly 'simple' Hartree-Fock theory, a so-called mean-field theory that is used for calculating approximately the quantum mechanical ground states of many-electron atoms, is nonlinear in the wave function. Furthermore, elements from classical mechanics and quantum mechanics are often mixed, for instance when describing electrons as balls when explaining the origin of the electric resistance, or when calculating the configuration of molecules by assuming that the atomic nuclei have well-defined positions in space. Whether on purpose or unintentionally, many courses and textbooks made us students believe that in principle everything follows from a set of fundamental laws, but that in practice it is unconvenient or unfeasable to do the exact calculation, and therefore approximations, plausible assumptions, intuitive models, and phenomenological arguments are made.  Only years later, I slowly began to understand that my problems had not primarily been due to a lack of talent and understanding on my side, but that there are indeed fundamental and interesting issues behind all these questions, some of which are the subject of lively discussions in the philosophy of physics. 

Probably the most important factor that made me change my views was becoming  a physics professor and teaching these courses myself. In particular the courses on statistical physics and condensed matter theory showed me that even the most advanced textbooks and research articles contain concepts, arguments, and steps that are a far cry from a strict derivation of the phenomena exclusively from a set of basic axioms or mathematically expressed laws. Looking back to my time as a student, I wish I had been taught about the interesting philosophical questions surrounding physics. Because we were lacking this information, I and probably many of my fellow students thought that physics is the most fundamental science, and that by learning quantum physics and particle physics we would learn the laws that rule nature at the most fundamental level.  

As a consequence of this experience, I adopted in the meantime the habit of pointing these philosophical issues out to students when presenting the course material or when teaching seminars. When writing down the deterministic, time-symmetric equations of classical mechanics, electrodynamics, or quantum mechanics, I address the question whether this implies that nature is indeed deterministic or time-symmetric. When mentioning probabilities in statistical mechanics, I address the question how this relates to the supposedly more 'fundamental', deterministic microscopic theories. When presenting the various models and methods of condensed-matter theory, I discuss how these models contain a mixture of elements from quantum and classical physics. Furthermore,  I now have discussed the issue of emergence and reduction for a couple of years in a seminar that I teach during the winter semester to master students of physics. 

The following pages will explain in a more detailed manner that condensed matter physics cannot be fully reduced to the supposedly 'fundamental' theory, which is quantum mechanics of $10^{23}$ particles. This means that many properties of condensed-matter systems are strongly emergent. It also means that the macroscopic properties of condensed matter systems have a top-down causal influence on their constitutents. In this way my contribution relates to the overall topic of this book and the workshop from which it results, which is top-down causation. 


 The outline of this paper is as follows: First, I will give a series of examples of condensed-matter systems that show emergent phenomena and that illustrate the issues to be discussed subsequently. Then, I will define the concepts of reduction and emergence, with a focus on the distinction between weak and strong emergence,  as they will be used later in the article. Next, using the texts by three Nobel laureates in condensed matter theory, I will show how condensed matter research is done in practice, and I will supplement it with insights from my own field of expertise, which is statistical physics. Based on all this information, we will then obtain list of reasons for accepting strong emergence in physics. Finally, I will deal with some widespread objections.

\section{Example systems}

Solids, liquids and gases are systems of $10^{23}$ particles that show many properties that the particles themselves don't possess: pressure, temperature, compressibility, electric conductivity, magnetism, specific heat, crystal structure, etc. It is an important goal of statistical physics and condensed matter theory to explain or even predict these properties in terms of the constituent atoms or molecules and their interactions. And these two fields of physics have indeed been very successful at relating these properties to the microscopic makeup of the respective system. 

One reason for this success is that these systems can be discussed without need to refer explicitly to their context. The listed properties are properties of an equilibrium system. The wider context enters only implicitly as it determines which objects are present, how they are arranged, and what are the environmental variables, such as temperature or pressure or the applied electrical or magnetic field.  



Other condensed matter systems are open systems or driven systems: they can show patterns and structures that depend crucially on their being embedded in a certain context, as they obtain a continuous input of energy and/or matter from their environment and pass energy and matter to their environment in a different form. An important example is thermal convection: when a gas or liquid is heated from below such that it is cooler at the upper surface than at the lower, the gas or liquid can be set into motion to form of convection rolls and thus transports heat efficiently from the warm to the cool surface. Such differences in temperature drive to a considerable part the weather and climate on earth.   In this case, the temperature differences are due to the sun's radiation heating the earth surface more at some places than at others. 
A particularly fascinating example of patterns in open systems are spiral patterns. They are for instance observed in the Belousov-Zhabotinsky reaction, which is an oscillating chemical reaction that cycles through three different reaction products, with the presence of one aiding the production of the next one. When put in a petri dish, one can observe spiral patterns. This pattern persists only for a limited time unless reaction products are continually removed and reaction educts are continually added. 
The most complex open systems are living systems. They require a continuous supply of resources from which they obtain their energy, for instance oxygen and food, and they pass the reaction products, such as carbon dioxide and water, to the environment. More importantly, they communicate continuously with their environment, obtaining from it cues about danger and food, and shaping it for instance by building burrows, farming lice, exchanging information with their conspecifics and other species, and by pursuing the production and survival of their offspring. 

Driven systems are also investigated by theoretical physicists, employing theories that refer to the parts of the systems and their interactions, however, such theories use a different level of description and don't usually refer to quantum mechanics. In this article, we will mainly focus on the first class of systems, as they appear to be less complex and more easily accessible to a description by a fundamental microscopic theory.

\section{Defining reduction and emergence}

As already indicated, the natural sciences, in particular physics, work in a reductionist way: they aim at explaining the properties of a system in terms of its constitutents and their properties, and they are very successful at this endeavor. More precisely, physicists perform a \emph{theoretical reduction} (and not an ontological reduction), which means that the property that one wants to explain is obtained mathematically by using elements of a microscopic theory that deals with the constituents of the system. Thus, for instance, the pressure of a gas is expressed in terms of the force that the atoms exert on the surface when they are reflected from it, and the temperature in terms of the kinetic energy of the atoms. The electrical resistivity is obtained from a mathematical description of the collisions of electrons with lattice vibrations and crystal defects. The oscillations and waves observed in the Belousov-Zhabotinsky reaction are reproduced from a set of chemical reaction equations for the concentrations of the molecules occurring in the system. Even beyond the realm of physics proper this type of approach is successful: the occurrence of traffic jams can be calculated based on models for cars that follow simple rules involving preferred velocities and preferred distances, the dynamics of ecosystems from biological populations feeding on each other, and stock market crashes from models of interacting agents.  

Emergence in a very general sense is the occurrence of properties in a system that the constituents don't have. In the previous section, examples for such emergent properties were given. 
So far there is nothing contentious about these general concepts of emergence and reduction. The discussion starts when one wants to assess the extent and nature of this reduction or emergence. Broadly speaking, there are two possible points of view: Either reduction is complete, at least in principle, and therefore emergence is weak; or reduction is incomplete and emergence is strong. Complete reduction means that the property or phenomenon to be described is completely contained in and implied by the microscopic theory. Incomplete reduction means that although the microscopic theory is successfully used, one invokes additional assumptions, approximations, hypotheses, or principles that are not part of the microscopic theory. Concordantly, weak emergence means that although the system has new properties, which are the emergent properties, these are fully accounted for by the microscopic theory. Strong emergence, in contrast, means that there is an irreducible, generically new aspect to the properties of the system as a whole that is not contained in or implied by the underlying theory. Strong emergence is closely associated with top-down causation. If higher-level properties are not fully accounted for by the lower-level theory, but if they affect the behavior of the system, then they also affect the way in which the constituents behave. This means that there is a top-down causation in addition to bottom-up causation.   

The picture underlying these concepts of emergence and reduction is that of a hierarchy of entities and systems, with the objects of one hierarchical level being the parts of the objects on the next level. Several such hierarchies can be constructed. One hierarchy that leads up to human societies is the following: Elementary particles -- atoms -- molecules -- cells -- individuals -- societies. Another hierachy that stays within the world of physics is this one: Elementary particles -- atoms -- solids and fluids -- planets and stars -- solar systems -- galaxies -- universe. 


In this article, we will mainly deal with condensed matter systems such as given in the first group of examples above. This means that we focus on the relation between the physics of mesoscopic or macroscopic systems and the relevant microscopic theory, which is a quantum mechanical description of approximately $10^{23}$ atomic nuclei and their surrounding electrons. In terms of the hierarchy of objects, we discuss the relation between atoms and solids. 

\section{Condensed matter research in practice}

In the following, I will quote three Nobel Laureates in condensed matter theory  as they explain how research in their field proceeds and how it is related to the supposedly fundamental quantum theory. All of them present strong arguments against a reductionist view. Nevertheless, two of them make surprisingly ambivalent statements concerning the nature of reduction. 


Since condensed-matter physics often uses methods from statistical mechanics and since statistical mechanics is my field of expertise, I will present additional arguments against the reductionist view by explaining how statistical mechanics relates to quantum mechanics and also to classical mechanics (which is for historical reasons still often used as a fundamental microscopic theory when justifying the laws of statistical mechanics).

\subsection{Philip Warren Anderson and the topic of symmetry breaking}

In his famous paper 'More is different' \cite{anderson1972more}, Anderson begins with the following sentences:  
\begin{quote}
The reductionist hypothesis may still be a topic for controversy among philosophers, but among the great majority of active scientists, I think it is accepted without question.  The workings of our minds and bodies, and of all the animate or inanimate matter of which we have any detailed knowlddge, are assumed to be controlled by the same set of fundamental laws.
\end{quote}
 Somewhat further on, he explains: 
\begin{quote}
(...) the reductionist hypothesis does not by any means imply a 'constructionist' one: The ability to reduce everything to simple fundamental laws does not imply the possibility to start from those laws and reconstruct the universe. (..) At each level of complexity entirely new properties appear, and the understanding of the new behaviors requires research which I think is as fundamental in its nature as any other.
\end{quote}
Then he gives the reason why this is so: It is because of symmetry breaking. The microscopic theory is invariant under spatial translation or rotation and thus for instance forbids the occurence of nonzero electrical dipole moments or of crystal lattices. This is because in an electrical dipole the negative charges have a different center than the positive charges, which means that a particular direction -- that in which the negative charges are shifted -- is singled out. Similarly, the atoms on a lattice have preferred positions, and the lattice has main axes that point in specific directions. When a direction is singled out, rotational symmetry, which treats all directions equally, is broken. When lattice positions are singled out, translational symmetry, which treats all positions equally, is broken. Another important case of symmetry breaking is the handedness of many biological molecules. This is an example that Anderson focuses on:  While very small molecules, such as ammonia, can tunnel between the two tetrahedral configurations with a high frequency, and thus have on average zero electrical dipole moment, larger molecules, such as sugar, cannot tunnel during any reasonable time period. Symmetry breaking is very important in condensed matter physics:  When the energetically favored state is a symmetry-broken state, the system chooses spontaneously one of these possible states, all of which have the same energy, as is analyzed in depth in the theory of phase transitions. This is relevant to phenomena such as superconductivity, magnetism, and superfluidity. Anderson's conclusion is that because of symmetry breaking the whole becomes different from the sum of its parts. 

There is an interesting dichotomy in this article: on the one hand, Anderson emphasizes more than once that he accepts reductionism (in the sense that 'everything obeys the same fundamental laws'), on the other hand he gives great arguments why reductionism does not work. Even more, he admits that there is a logical contradiction between the laws of quantum mechanics and the fact the molecules have symmetry-breaking structures. Since he wrote his article, there has been a lot of progress in this field, and several authors have pointed out that in fact the problems of symmetry breaking and of molecular structure are closely related to the problems of interpreting the quantum mechanical measurement process and of understanding the relation between quantum mechanics and classical mechanics \cite{primas2013chemistry,chibbaro2014reductionism,matyus2018pre}. We will address this problem further below.  

\subsection{Robert Laughlin and higher-order principles} 

In his often-cited article with Pines ``The Theory of Everything'' \cite{laughlin2000theory}, and also in his popular-science book ``A different universe'' \cite{laughlin2008different} Laughlin expresses a similar view to Anderson. The authors of \cite{laughlin2000theory} label the Schr\"odinger equation that describes all the electrons and nuclei of a system as the ``Theory of Everything'' for condensed-matter physics and explain:
\begin{quote} 
We know that this equation is correct because it has been solved accurately for small numbers of particles and found to agree in minute detail with experiment. However, it cannot be solved accurately when the number of particles exceeds about 10. No computer (...) that will ever exist can break this barrier because there is a catastrophy of dimension.  (....) We have succeeded in reducing all of ordinary physical behavior to a simple, correct Theory of Everything only to discover that it has revealed exactly nothing about many things of great importance.\end{quote}
Just as Anderson, the authors first appear to commit to reductionism, and then explain why it is of no use for many important problems in condensed matter theory. Confusingly, they do not clarify what they mean by ``correct'' or ``accurate'';  they cannot possibly mean that the theory is exact as they write themselves that it neglects several effects (such as the coupling to photons). Apparently they think that these effects are unimportant even when dealing with solids consisting of $10^{23}$ atoms. 

Their arguments why the ``Theory of Everything'' is useless for condensed matter theory include symmetry breaking, but are more general: In his book, Laughlin calls the principles determining many emergent features 'Higher-order principles'  (HOPs). In the article, he and Pines explain it as follows:   
\begin{quote}
Experiments of this kind (i.e., condensed-matter experiments that allow us to measure the natural constants with extremely high precision) work because there are higher organizing principles in nature that make them work. The Josephson quantum\footnote{The Josephson quantum is due to the fact that magnetic flux surrounded by a superconducting current can only be an integer multiple of a basic flux unit. This is because the magnetic flux affects the phase of the superconducting wave function, but the wave function must have a again the same phase after going around the flux line once.} is exact because of the principle of continuous symmetry breaking. The quantum Hall effect \footnote{The Hall effect gives rise to a transverse electric field in a two-dimensional conductor to which a magnetic field is applied that is perpendicular to the current. The Hall resistance is the ratio of the transverse electrical field to the electrical current, and it takes quantized values when the material is cooled down sufficiently. The explanation of this effect involves considerations about the topology of the wave function.} is exact because of localization. (....) Both are transcendent, in that they would continue to be true and to lead to exact results even if the Theory of Everything was changed.
\end{quote}
 They go on to explain that ``for at least some fundamental things in nature the Theory of Everything is irrelevant.'' They introduce the concept of a quantum protectorate, 
\begin{quote}
a stable state of matter whose generic low-energy properties are determined by a higher organizing principle and nothing else. (...) The low-energy excited quantum states of these systems are particles in exactly the same sense that the electron in the vacuum of quantum electrodynamics is a particle. (...) The nature of the underlying theory is unknowable until one raises the energy scale sufficiently to escape protection.
\end{quote}
In fact, even though they do not mention this,  the Theory of Everything, i.e., the many-particle Schr\"odinger equation,  is also such a quantum protectorate, valid at low energies, obtained from general considerations, and independent of more microscopic theories, such as quantum field theories or string theories. 

All this raises the question whether it is logically consistent to claim that a system is determined by a micropscopic law and to claim simultaneously that this law is irrelevant since higher-order principles govern the behavior of the system. This is in my view the most interesting question raised by the article, but the authors don't address it. We will come back to this question later. While the examples chosen by Anderson were taken mainly from the room-temperature classical world, the examples used by Laughlin and Pines are mainly quantum phenomena. Therefore, the issue now is not primarily the relation between the quantum description and the classical description, but the relation between different quantum descriptions (the microscopic one and the one based on HOPs) of the same system. 

\subsection{Anthony Leggett and the quantum-classical transition}

In his article ``On the nature of research in condensed-state physics'' \cite{leggett1992nature}, Leggett writes
\begin{quote}
No significant advance in the theory of matter in bulk has ever come about through derivation from microscopic principles. (...) I would confidently argue further that it is in principle
and forever impossible to carry out such a derivation. (...) The so-called derivations of the results of solid state
physics from microscopic principles alone are almost all bogus, if
'derivation' is meant to have anything like its usual sense.
\end{quote}
So he agrees with Anderson and Laughlin that the microscopic theory is useless for deriving the properties of solids. He illustrates this by using the example of Ohm's law, i.e.~the law for the electrical resistance: 
\begin{quote}
Consider as
elementary a principle as Ohm's law. As far as I know, no-one has ever
come even remotely within reach of deriving Ohm's law from microscopic
principles without a whole host of auxiliary assumptions ('physical
approximations'), which one almost certainly would not have thought of
making unless one knew in advance the result one wanted to get, (and
some of which may be regarded as essentially begging the question). This
situation is fairly typical: once you have reason to believe that a certain
kind of model or theory will actually work at the macroscopic or intermediate
level, then it is sometimes possible to show that you can 'derive'
it from microscopic theory, in the sense that you may be able to find the
auxiliary assumptions or approximations you have to make to lead to the
result you want.
\end{quote}
But if auxiliary assumptions and approximations are involved in the derivation of Ohm's law, the theory or even the phenomenon might not be contained in the microscopic theory. This becomes even clearer in the next paragraph:
\begin{quote}
 But you can practically never justify these auxiliary
assumptions, and the whole process is highly dangerous anyway: very often
you find that what you thought you had 'proved' comes unstuck
experimentally (for instance, you 'prove' Ohm's law quite generally only
to discover that superconductors don't obey it) and when you go back to
your proof you discover as often as not that you had implicitly slipped in
an assumption that begs the whole question. (...)
\end{quote}
This experience makes him skeptical with respect to the validity of his calculations:
\begin{quote}
Incidentally, as a psychological fact, it does occasionally happen that one
is led to a new model by a microscopic calculation. But in that case one
certainly doesn't believe the model because of the calculation: on the contrary,
in my experience at least one disbelieves or distrusts the calculation
unless and until one has a flash of insight and sees the result in terms of
a model.
\end{quote}
Now comes the most interesting part about the nature of theories in condensed-matter physics:
\begin{quote}
I claim then that the important advances in macroscopic physics come
essentially in the construction of models at an intermediate or macroscopic
level, and that these are logically (and psychologically) independent of
microscopic physics. Examples of the kind of models I have in mind which
may be familiar to some readers include the Debye model of a crystalline
solid, the idea of a quasiparticle, the Ising or Heisenberg picture of a
magnetic material, the two-fluid model of liquid helium, London's
approach to superconductivity (...) In some cases these models may be
plausibly represented as 'based on' microscopic physics, in the sense that
they can be described as making assumptions about microscopic entities
(e.g. 'the atoms are arranged in a regular lattice'), but in other cases (such
as the two-fluid model) they are independent even in this sense. What all
have in common is that they can serve as some kind of concrete picture,
or metaphor, which can guide our thinking about the subject in question.
And they guide it in their own right, and not because they are a sort of
crude shorthand for some underlying mathematics derived from 'basic
principles.' 
\end{quote}
This is reminiscent of the higher-order principles mentioned by Laughlin, but is more general: One needs to identify the important features that are responsible for the phenomenon to be explained. These features are discovered by an intuitive understanding of the system, often in terms of pictures and metaphors. They form the basis of a model, which in turn is the basis of an effective mathematical description. Further on, he explains that these models in fact often are incompatible with the microscopic theory:
\begin{quote}
(...) not only is there no good reason to believe that \emph{all} the properties of condensed-matter systems are simply consequences of the properties of their atomic-level constituents, but that \emph{there is a very good reason not to believe it.} (...) Indeed, I would be perfectly happy to share the conventional reductionist prejudice were it not for a single fact (...) which is so overwhelming in its implications that it forces us to challenge even what we might think of as the most basic common sense. This fact is the existence of the quantum measurement paradox.(...) this argument implies that quantum mechanics, of its nature, cannot give a \emph{complete} account of all the properties of the world at the macroscopic level. (...) It follows that somewhere along the line from the atom to human consciousness quantum mechanics must break down.
\end{quote}
I fully agree with his characterization of how condensed matter physics is done in practice and with his diagnosis that there must be limits to the validity of quantum mechanics. In fact, together with George Ellis, I have written a paper that interprets the quantum measurement process in terms of top-down effects from the classical world on the quantum world \cite{drossel2018contextual}. 



\subsection{Statistical physics and the concept of probabilities}

After these three examples from solid-state physics, I want to conclude this section with an example from my own field, which is statistical mechanics. Statistical mechanics does not deal with a particular system, but provides a framework for calculating properties of systems of $10^{23}$ particles at finite temperature, be they solids, liquids, or gases. Condensed matter theory often uses concepts and calculations taken from statistical mechanics, for instance when calculating the specific heat of a system, the phase transition between a paramagnet and a ferromagnet,  or the conditions (temperature, magnetic field, or size of electrical current) that destroy superconductivity. This means that one does not use the supposedly 'fundamental' theory but a theory that brings with it new concepts. 

The basic concept of statistical mechanics is that of the probability of a state of a system or a subpart of the system. For instance, a fundamental theorem of statistical mechanics, from which almost everything else can be derived, is that in an isolated system in equilibrium all states that are accessible to a system occur with the same probability. Such an equilibrium state has maximum entropy, and therefore this theorem is closely related to the second law of thermodynamics, which states that entropy increases in a closed system until the equilibrium state is reached. Part of the textbooks of statistical mechanics and articles of the foundations of statistical physics aim at deriving these probabilistic rules from a microscopic deterministic theory, either classical mechanics (where the atoms of a gas are treated as little hard balls) or quantum mechanics. A close look at these derivations, however, reveals that they in fact always put in by hand what they want to get out: randomness. The initial state of the system must be a 'random' state among all those that cannot be distinguished from each other when specifying them only with finite precision. The apparent randomness in the future time evolution then follows from this hidden, unperceivable initial randomness. This means that the randomness of statistical mechanics is not reduced to a deterministic 'fundamental' theory, but it is only moved back to the initial conditions and hidden there. Of course, a truly deterministic world would not leave us the freedom to say that the precise initial state does not matter because the real one can be any of them. The initial state would be the state it is and not care about our ingnorance of finer details. It is amazing that nature conforms to our ignorance and behaves as if there was nothing special to the initial state that would later lead to surprising effects in the time evolution. The natural conclusion, which however goes against the reductionist agenda, woud be to say that the mathematical concept of infinite precision underlying the equations of the 'fundamental' theories has no support from empirical science. In fact, when there are no specific top-down effects since the system is isolated, the system becomes maximally indifferent as to the state in which it tends to be. This is what the second law of thermodynamics states. This topic is discussed in more detail in publications by Nicolas Gisin \cite{gisin2016,gisin2018indeterminism} and by myself \cite{drossel2015relation,drossel2017ten}.

Interestingly, there is a close relation between the quantum measurement problem and the problem of relating statistical mechanics to quantum mechanics: quantum measurement and statistical mechanics both involve probabilities and irreversibility.  Both deal with systems of a macroscopic number of particles (in the case of quantum measurement this is the measurement apparatus). Concordantly, the theories dealing with the two problems involve similar ideas and types of calculations: they are based on decoherence theory which explains why quantum mechanical superpositions vanish when a quantum particle interacts with a macroscopic number of other particles. These calculations -- not surprisingly -- again need to invoke random initial states. In contrast to theories that 'derive' statistical mechanics from classical mechanics, these random initial states are however not sufficient in order to 'derive' statistical mechanics from quantum mechanics, because the calculations give all possible stochastic trajectories simultaneously instead of picking one of them for each realization. So we arrive again at the point emphasized by A. Leggett: Quantum mechanics is not consistent with physics of macroscopic, finite-temperature objects. This means that there must be limits of validity to quantum mechanics.

\section{Strong arguments for strong emergence in physics}

\subsection{A full reduction to a microscopic theory cannot be done}

The foregoing quotations and discussions have made clear that a full reduction is never done in practice, and that it is impossible for several reasons. The more trivial reason is limited computing power, not just at the present time but for all future since the calculation of the time evolution of the quantum state of as few as 1000 particles would require more information than contained in the universe. This means that the belief in full reduction is a metaphysical belief, as it can never, even in principle be tested. In contrast, physics is an empirical science rooted in what can be measured and observed. 

But beyond this, I think there are good reasons why reductionism even when taken as a metaphysical belief is wrong. As most clearly stated by A.~Leggett, there is a logical incompatibility between quantum mechanics, which leads to superposition of macroscopic objects being in different locations, and the observation that macroscopic objects are localized in space. 

It is my impression that in fact all the effective theories and models and approximations made in order to obtain the properties of macroscopic systems, involve assumptions and steps that are  in contradiction with the supposedly fundamental theory.  For instance, the derivations of the phonon (lattice vibration) spectrum of solids starts by separating the equation for the electrons from that for the nuclei by using the so-called Born-Oppenheimer approximation. This approximation is in fact a mixture of quantum theory and classical theory, as it assumes that the atomic nuclei are localized in space and not in superpositions of different locations. It is also the basis of quantum chemistry and the widely-used density functional theory.  A nice discussion of this can be found in the book \cite{chibbaro2014reductionism}. Similarly, other theories in condensed matter make assumptions and approximations that deviate from the linear, deterministic, microscopic Theory of Everything. It appears to me that they all put in contextual information that is not intrinsic to the Theory of Everything and that contains elements from classical physics and oftentimes also of statistical physics. 

And there are many more reasons why full reductionism is wrong, as listed in the following subsections.

\subsection{The parts have never existed without the whole}

The mental picture that many people have when they think of emergence is that there are first the parts, and then they get together and form the whole. But this picture is only half the truth. These parts would not be there without a larger context that permits their existence and determines their properties. Which types of elementary particles exist in the universe depends on the properties of the quantum vacuum, namely its symmetries and degrees of freedom. Whether these particles can combine to create larger objects, is again determined by the context: In the early universe, the temperature and density of the universe determined which types of objects existed in it: quarks and gluons, or nuclei and electrons, or atoms and radiation, or stars and galaxies and planets. 

On smaller scales, one observes the same extent of context dependence, and many examples are listed in the book by G.~Ellis \cite{EllisTopDown}:
The lifetime of a neutron depends on whether it is a free particle or part of a nucleus. In a crystal, the presence of the crystal structure permits the existence of phonons, and the symmetries of the crystal determine their properties. The mass of an electron depends on the band structure of the metal in which is is. Whether a substance is liquid or solid depends on the environmental conditions, such as temperature and pressure.  


\subsection{The laws of physics are not exact}

Full reductionism could only be correct if the supposedly fundamental laws were extremely accurate. Otherwise, even minute imprecisions could become magnified in macroscopic system to the extent that the fundamental theory cannot predict correctly the properties of the macroscopic system. But we know that our present theories are not fully exact: They are idealizations that leave aside many influences that affect a system. Newtonian mechanics, for instance, ignores the effect of friction, or includes it in a simple way, which is neither exact nor derived from a microscopic theory. A large part of thermodynamics is based on local equilibrium, which is not an exact but only an approximate description. Quantum field theory is burdened with exactly the same problems as nonrelativistic quantum mechanics, namely the discrepancy between a unitary, deterministic time evolution applied after preparation of the initial state and before measurement of the final state, and the probabilities and nonlinear expressions featuring in calculations of cross sections and transition rates.  Furthermore, it could not yet be harmonized with Einstein's theory of general relativity, which describes gravity.  

Batterman argues convincingly that physics theories are asymptotic theories that become exact only in an asymptotic limit where a quantity goes to zero or infinity \cite{batterman2001devil}. Newtonian mechanics is a good example of how the applicability of a theory depends on certain quantities being small or large: The velocity of light must be large compared to the velocities of the considered objects (otherwise one needs to use the theory of special relativity), energies must be so large that quantum effects can be ignored (otherwise one needs quantum mechanics), and distances must be small compared to cosmic distances on which the curvature of space is felt (otherwise one needs general relativity). In earlier times, there was a widespread belief that Newtonian mechanics is an exact theory, ony to realize in the last century that on all those scales that could not be explored before (such as the very small, the very large, and the very fast) Newtonian mechanics becomes invalid, and that it is a very good approximation, but not exact, on those scales that had been explored. We can expect that our present theories will also turn out to become inappropriate when new parameter ranges, which we could not explore previously, become accessible to experiments. 
In fact, it appears impossible to have exact, comprehensive microscopic laws that govern everything that happens, and to have at the same time a complete insensitiveness to these microscopic laws in systems that are determined by  higher-order principles.

\subsection{The microscopic world is not deterministic}

One of the main shocks caused by quantum mechanics is the insight that the microscopic nature is fundamentally indeterministic, as for instance visible in radioactive decay where nothing in the state of an atom allows one to predict when it will decay. Only the half life can be known and  be calculated from a microscopic theory. The same hold for the quantum-mechanical measurement process, where the experiment gives one of the possible outcomes with a probability that can be calculated using the rules of quantum mechanics, but the process itself is stochastic with nothing in the initial state determining which of the outcomes will be observed.

In order for full reductionism to hold, the microscopic theory must be deterministic. Only then does the microscopic theory determine everything that happens. Otherwise the microscopic theory can at best give probabilities for the different possible events. When dealing with a system of many particles, one likely outcome of such stochastic dynamics is ergodicity, where the system goes to an equilibrium state that is characterized by a stationary probability distribution. This is what happens for instance when a thermodynamic system such as a gas reaches its equilibrium state, which has  maximum entropy. The macroscopic state variables characterizing this equilibrium state, such as volume or pressure, are independent of the initial microscopic state and are essentially determined by the constraints imposed from the outside. Thermodynamic systems are in fact a nice example for top-down causation, as they do essentially nothing else but to adjust to whatever state is imposed on them by manipulations from the outside, such as volume change or energy input. 

But even when chance does not lead to ergodicity, top-down causation is involved. When a quantum measurement event happens, the measurement apparatus determines the possible types of measurement outcomes (for instance whether the z component or the x component of the spin is measured). When an excited atom goes to its ground state by emitting a photon, it can do so only because the surrounding medium, the quantum vacuum, can take up that photon. When the atom is enclosed in a small cavity the size of which is not a multiple of the photon wavelength, the photon cannot be emitted. It thus appears that all instances of quantum chance are in fact strongly dependent on top-down causation. A quantum object by itself, when it is carefully isolated from interacting with the rest of the world, evolves according to the 'fundamental', deterministic, and linear quantum mechanical equations.  

Karl Popper made the interesting suggestion that chance at the lower level is necessary for top-down causation from the higher level \cite{Popper1977-POPTSA}, and he has been criticized for it \cite{sep-properties-emergent}. However, I think that he is right. Only when the entities at the lower level are not fully controlled by the microscopic laws can they respond to the higher level. It is often argued that random changes are as little susceptible to top-down effects as are deterministic changes. But this is based on the wrong premise that stochasticity is an intrinsic property of the lower-level system, while it arises in fact from the interaction of the lower-level constituents with the larger context. 

\subsection{Emergent properties are insensitive to microscopic details}

As mentioned above in Sec.~4.2, Laughlin has emphasized a lot that emergent properties are insensitive to microscopic details. It is the higher-order principles that determine the behavior of the systems he discusses. There are indeed many examples where systems that are microscopically different show the same macroscopic behavior and are described by the same mathematical theory. For instance, the phase transition from a paramagnet to a ferromagnet below the Curie temperature is described by the same mathematics as the Higgs mechanism that can successfully explain how particles obtain  their mass. Furthermore, within one type of systems, for instance a ferromagnet made of iron, the mathematical description in the form of macroscopic or effective variables is independent of the precise microscopic arrangement of atoms and defects. There exist many microscopic realizations of the same macroscopic state. This is also emphasized in the book by George Ellis when he mentions the concept of  'multiple realizability' of a macroscopic state by microscopic states. When additionally the time evolution can be expressed in terms of the macroscopic variables alone, this means that the causal processes can be described on the level of the macrostate. A natural conclusion is that in such cases causal processes do indeed occur at the level of the macrostates and that the microstates merely adjust to the constraints imposed by the macrostate.


\subsection{Many systems are inseparable from their environment}

So far, we have mainly addressed equilibrium systems, which can be cut off from their environment without losing their properties. In fact, this is an approximate statement, as no system can be completely cut off from the environment; the best one can do is putting a system into a closed box or lab with no directed input or output of energy or matter. The typical solid-state systems discussed by the above-cited Nobel Laureates remain what they are under such conditions, at least on time scales relevant for experiments. On the very long run, they will be changed by surface reactions, by radioactive radiation, and ultimately by the sun turning into a red giant and burning everything on earth. In contrast to these relatively stable systems, open (or dissipative) systems require an ongoing input and output of energy and/or matter in order to remain what they are. We have given in Section 2 the examples of convection patterns, oscillating reactions, and living organisms. For these systems, the idea that they are determined by their parts and the interactions of their parts, is completely wrong, as they are what they are only in contact with their environment: Convection patterns require an input of heat at one end and a cooling surface at the other end; oscillating reaction can only be sustained when certain reactants are continually supplied and certain reaction products removed; living beings need to breathe and feed, and they respond in a complex way to cues from their environment. Since all these systems exist only due to being continually sustained by their environment, it is completely wrong to think of them merely in terms of their parts and the interactions of their parts. 
The emergent features of these systems are therefore clear-cut cases of top-down causation.

\section{Answers to objections}

When discussing the issue of strong emergence in physics, a variety of objections are being made, which shall be dealt with in the following.

\subsection{We will find a more fundamental theory}

Quite a few scientists hope that even if our present microscopic theories are not yet the final, correct theories, the progress of physics will ultimately lead to such final theories. 

If such theories will ever be found, they must achieve a lot: They must solve the quantum measurement problem, and they must also establish a relation between general relativity and quantum physics. In the light of the arguments presented in this article, it seems impossbile that such a theory exists. The top-down effects of the larger, macroscopic context on the microscopic constituents cannot be captured by a purely microscopic theory. 

Some people argue that even if the ultimate theory might be unknowable, nature cans nevertheless be governed by basic laws which are known to us only in approximate versions. This is a valid philosophical position, but it appears to be rooted more in a metaphysical commitment than in emiprical evidence. This commitment is to physicalism and to causal closure. Both are very restrictive assumptions that can be doubted on philosophical grounds. The strongest arguments against physicalism are based on human consciousness, and are brought forward by authors such as Brigitte Falkenburg \cite{falkenburg2012mythos}, Thomas Nagel \cite{nagel2012mind}, or Markus Grbriel \cite{gabriel2017not}. All the arguments brought forward by George Ellis in favor of top-down causation are also arguments against causal closure. These arguments show that material systems are susceptible even to non-material influences such as goals, ideas, or man-made conventions.

\subsection{You argue from our lack of knowledge, this is dangerous}

In the 19th and early 20th centure, the British emergentists argued in favor of strong emergence based on chemistry and biology \cite{mclaughlin1992rise,sep-properties-emergent}. They could not imagine that chemistry of biology obeys the laws of physics, and they assumed that complex objects are subject to different laws. But after the successes of quantum chemistry at explaining the periodic table and calculating molecular structures, and the success of biology at reducing inheritance to the properties of the DNA molecule, these emergentist positions came in discredit. By analogy, those who hold to strong emergentist views, are told that progress of science will probably prove them wrong. If we cannot explain a macroscopic feature in terms of a microscopic theory today, it might become possible in the future. 

However, the arguments presented in this article are not arguments from ignorance, but from the very nature of the systems under study. Even the claimed reduction of chemistry or biology to (quantum) physics is incomplete. It is a partial reduction that invokes a collection of models and arguments that includes features from the quantum world as well as from the classical world, as explained in Sec.~5.1. In particular life is so dependent on its environment that the reductionist enterprise is doomed to failure on principal grounds.

\subsection{There are fully reductionist explanations for the quantum-classical transition and the second law of thermodynamics}

This is an expert discussion that goes into the details of the theories offered. As I have tried to convey above, all these theories need to invoke additional concepts beyond the 'fundamental' Theory of Everything. In one form or another, they all rely on some type of randomness of initial states or environmental states. Furtheremore, since quantum mechanics is a linear theory, one cannot avoid the resulting superpositions of macroscopic states. Pointing out that these superpositions can look like a classical combination of the different possible outcomes with their associated probabilities does not fully solve the problem, as nature realizes in each instance only one of the possible outcomes. Of course, there are interpretations of quantum mechanics that deal with this issue (such as the many-worlds interpretation, the statistical interpretation, consistent histories, relational interpretation), but all of them give up on the goal of science of accounting for an objective, observer-independent reality with its contingent, particular history.  I am not willing to abandon this goal, and I think that abandoning this goals hampers the progress of science and distracts from the open problems that await solutions.

\subsection{
All supposed top-down effects can equally well be expressed in terms of a microscopic theory}

The idea behind this objection is that the context that produces the top-down effects is itself composed of atoms and can be described by a microscopic theory. Instead of having the description in terms of a system and a context, one could describe both system and context microscopically. 

However, apart from being impossible in practice, there are several reasons why this is not possible in principle: First, there is no isolated system, since every system emits thermal radiation to the environment, which in turn passes it on to open space. Furthermore, every system is exposed to the influence of gravitational forces, which cannot be shielded by any means. Third, having a closed system leads again to all the problems related to interpreting quantum mechanics. 

Some authors hold that the universe as a whole is a closed system and can therefore be described by a microscopic theory for all its particles and their interactions. However, the driving force behind everything that happens in the universe is the expansion of the universe, starting from a very special initial state. Neither the initial state, nor the expansion results from the interaction of the particles, and therefore the claim that the universe is determined by its parts and their interactions is wrong. 

\section{Conclusion}

To conclude, I see many reasons to reject the view that the world of physics is causally closed with everything being determined bottom-up by fundamental microscopic laws. As stated in the book by George Ellis: The lower (microscopic) level enables everything and  underlies everything, but does not determine everything. The higher hierarchical levels have an important say at what happens in nature. 
Anthony Leggett writes in his above-cited article that the non-reductionist view is a minority view among professional physicists.  In fact, I am not so sure about this. Clearly, the majority of people who write and talk on such foundational topics, have a more-or-less reductionist view. However, when I talk to colleagues it appears to me that many of them are aware that the world described by physics is not a monolithic block that is controled by a small set of rules. Maybe that those people who  have not simple clear view but think that physics is more complex  usually don't write papers on this topic. This is one of the reasons why I decided to write this paper. 


\begin{thebibliography}{10}
\bibitem{anderson1972more}
Philip~W Anderson.
\newblock More is different.
\newblock {\em Science}, 177(4047):393--396, 1972.

\bibitem{primas2013chemistry}
Hans Primas.
\newblock {\em Chemistry, quantum mechanics and reductionism: perspectives in
  theoretical chemistry}, volume~24.
\newblock Springer-Verlag Heidelberg Berlin, 2013.

\bibitem{chibbaro2014reductionism}
Sergio Chibbaro, Lamberto Rondoni, and Angelo Vulpiani.
\newblock {\em Reductionism, emergence and levels of reality, Ch. 6}.
\newblock Springer, 2014.

\bibitem{matyus2018pre}
Edit Matyus.
\newblock Pre-born-oppenheimer molecular structure theory.
\newblock {\em arXiv preprint arXiv:1801.05885}, 2018.

\bibitem{laughlin2000theory}
Robert~B Laughlin and David Pines.
\newblock The theory of everything.
\newblock {\em Proceedings of the National Academy of Sciences of the United
  States of America}, pages 28--31, 2000.

\bibitem{laughlin2008different}
Robert~B Laughlin.
\newblock {\em A different universe: Reinventing physics from the bottom down}.
\newblock Basic books, 2008.

\bibitem{leggett1992nature}
Anthony~J Leggett.
\newblock On the nature of research in condensed-state physics.
\newblock {\em Foundations of Physics}, 22(2):221--233, 1992.

\bibitem{drossel2018contextual}
Barbara Drossel and George Ellis.
\newblock Contextual wavefunction collapse: An integrated theory of quantum
  measurement.
\newblock {\em New Journal of Physics}, 20(11):113025, 2018.

\bibitem{gisin2016}
Nicolas Gisin.
\newblock Time really passes, science can’t deny that.
\newblock In Stupar~S. Renner~R., editor, {\em Time in Physics}, pages 1--15.
  Birkhauser, 2017.

\bibitem{gisin2018indeterminism}
Nicolas Gisin.
\newblock Indeterminism in physics, classical chaos and bohmian mechanics. are
  real numbers really real?
\newblock {\em arXiv preprint arXiv:1803.06824}, 2018.

\bibitem{drossel2015relation}
Barbara Drossel.
\newblock On the relation between the second law of thermodynamics and
  classical and quantum mechanics.
\newblock In {\em Why More Is Different}, pages 41--54. Springer, 2015.

\bibitem{drossel2017ten}
Barbara Drossel.
\newblock Ten reasons why a thermalized system cannot be described by a
  many-particle wave function.
\newblock {\em Studies in History and Philosophy of Science Part B: Studies in
  History and Philosophy of Modern Physics}, 58:12--21, 2017.

\bibitem{EllisTopDown}
George Ellis.
\newblock {\em {\it How can physics underlie the mind?: top-down causation in
  the human context}}.
\newblock Springer Heidelberg, 2016.

\bibitem{batterman2001devil}
Robert~W Batterman.
\newblock {\em The devil in the details: Asymptotic reasoning in explanation,
  reduction, and emergence}.
\newblock Oxford University Press, 2001.

\bibitem{Popper1977-POPTSA}
Karl~R. Popper and John~C. Eccles.
\newblock {\em The Self and Its Brain: An Argument for Interactionism}.
\newblock Springer Heidelberg Berlin, 1977.

\bibitem{sep-properties-emergent}
Timothy O'Connor and Hong~Yu Wong.
\newblock Emergent properties.
\newblock In Edward~N. Zalta, editor, {\em The Stanford Encyclopedia of
  Philosophy}. Metaphysics Research Lab, Stanford University, summer 2015
  edition, 2015.

\bibitem{falkenburg2012mythos}
Brigitte Falkenburg.
\newblock {\em Mythos Determinismus: wieviel erkl{\"a}rt uns die
  Hirnforschung?}
\newblock Springer-Verlag Berlin Heidelberg, 2012.

\bibitem{nagel2012mind}
Thomas Nagel.
\newblock {\em Mind and cosmos: why the materialist neo-Darwinian conception of
  nature is almost certainly false}.
\newblock Oxford University Press, 2012.

\bibitem{gabriel2017not}
Markus Gabriel.
\newblock {\em I am Not a Brain: Philosophy of Mind for the 21st Century}.
\newblock John Wiley \& Sons, 2017.

\bibitem{mclaughlin1992rise}
Brian McLaughlin et~al.
\newblock The rise and fall of british emergentism.
\newblock {\em Emergence or reduction}, pages 49--93, 1992.

\end{thebibliography}
\bibliographystyle{unsrt}

\end{document}